\documentclass[twocolumn,showpacs,%
  nofootinbib,aps,superscriptaddress,%
  eqsecnum,prd,notitlepage,showkeys,10pt]{revtex4-1}

\usepackage{amssymb}
\usepackage{amsmath}
\usepackage{graphicx}
\usepackage{dcolumn}
\usepackage{hyperref}
\usepackage{amsfonts}
\usepackage{lipsum} % just for demonstration
\usepackage{fancybox}
\usepackage{adjustbox}
\usepackage{caption}
\usepackage{float}
\usepackage{pgfplots}
\usepackage{url}
\usepackage{algorithm}
\usepackage{graphicx}
\usepackage{algpseudocode}

\newcommand\tab[1][1cm]{\hspace*{#1}}
\newcommand{\xdownarrow}[1]{%
  {\left\downarrow\vbox to #1{}\right.\kern-\nulldelimiterspace}
}

\makeatletter
\newcommand*{\cleartoleftpage}{%
  \clearpage
    \if@twoside
    \ifodd\c@page
      \hbox{}\newpage
      \if@twocolumn
        \hbox{}\newpage
      \fi
    \fi
  \fi
}
\makeatother

\begin{document}
\title{\LARGE{NtMalDetect: A Machine Learning Approach to Malware Detection Using Native API System Calls}}
\author{Chan Woo Kim (chanwkim01@gmail.com)}
\affiliation{Shanghai American School Puxi}

\begin{abstract}
\vspace*{0.3cm}

	\begin{center}
		\emph{\bf{Abstract}} 
	\end{center}
\tab As computing systems become increasingly advanced and as users increasingly engage themselves in technology, security has never been a greater concern. In malware detection, static analysis, the method of analyzing potentially malicious files, has been the prominent approach. This approach, however, quickly falls short as malicious programs become more advanced and adopt the capabilities of obfuscating its binaries to execute the same malicious functions, making static analysis extremely difficult for newer variants. The approach assessed in this paper is a novel dynamic malware analysis method, which may generalize better than static analysis to newer variants. Inspired by recent successes in Natural Language Processing (NLP), widely used document classification techniques were assessed in detecting malware by doing such analysis on system calls, which contain useful information about the operation of a program as requests that the program makes of the kernel. Features considered are extracted from system call traces of benign and malicious programs, and the task to classify these traces is treated as a binary document classification task of system call traces. The system call traces were processed to remove the parameters to only leave the system call function names. The features were grouped into various n-grams and weighted with Term Frequency-Inverse Document Frequency. This paper shows that Linear Support Vector Machines (SVM) optimized by Stochastic Gradient Descent and the traditional Coordinate Descent on the Wolfe Dual form of the SVM are effective in this approach, achieving a highest of 96\% accuracy with 95\% recall score. Additional contributions include the identification of significant system call sequences that could be avenues for further research.

\end{abstract}

\maketitle

\section {Introduction}
Static malware analysis has been the prominent approach in malware detection. Static analysis pertains to analyzing binaries of programs without executing them. Although many valid approaches have been proposed, research suggests that the binary code obfuscation techniques that are available today are incredibly difficult to overcome \cite{moser2007limits}. For the same reason, although static analysis can often accurately detect known malware, it struggles against new variants and zero-day threats \cite{canzanese2015detection}. Regarding this, an approach that may resolve this issue is dynamically analyzing a given program to detect if it is malicious, thereby analyzing the behavior of programs instead. The idea is that even if a malicious file changes, its behavior should remain the same. Dynamic analysis aims to find patterns in program execution, training a program to do which will allow future detection of malicious behavior, regardless of its code structure or whether or not it has been found before.

Some approaches to dynamic analysis of malware include looking for files that have been added or modified, newly installed services, newly running processes, registry modifications, and more \cite{distler2007malware}. One such method is analyses of system calls. System calls are routines user programs call to use services of the operating system \cite{hubballi2011sequencegram}. Any program running in an operating system has a definitive set of system calls. By analyzing sequences of system calls of programs running in normal operating conditions, one may gain insight in the abnormality of processes executed by a given malicious program by analyzing to what extent it diverges from usual behavior. FIG. 1 shows an example Windows Native API system call trace.

\begin{figure}

\fbox{
    \parbox{24em}{
    \texttt{
    \begin{flushleft}
        NtQueryPerformanceCounter( Counter=0xbcf6c8 [1.45779e+009], Freq=null ) => 0 \newline
NtProtectVirtualMemory( ProcessHandle=-1, BaseAddress=0xbcf6f4 ?\newline
NtProtectVirtualMemory( ProcessHandle=-1, BaseAddress=0xbcf6f4 [0x7702e000]?\newline
NtQuerySystemInformation( SystemInformationClass=0 [SystemBasicInformation]\newline
NtQueryVirtualMemory( ProcessHandle=-1, BaseAddress=0x76f20000,
\end{flushleft}
  }
  }
}   

\caption{\label{fig:your-figure}Example System Call Trace (truncated to five calls) }
\end{figure}

The task to classify traces of system calls as belonging to either a malicious or a benign program is treated in parallel to a binary document classification task. Document classification is a form of machine learning, a subset of Natural Language Processing (NLP), aiming to assign a category or a class to a document by analyzing its content \cite{ghaffari}. With recent improvements and successes in document classification, it was deemed appropriate to utilize its methods to evaluate its effectiveness in this regard \cite{metz_2017}. 
	The purpose of this research is to evaluate the effectiveness of a machine learning approach in malware detection, using document classification techniques to conduct dynamic analysis of malicious programs using system call traces. A trace of system calls of a given program will be equivalent of a document in a document classification task, and by training classifiers on the dataset, we aim to classify previously undetected malicious programs by making predictions based on their system calls.

\section {Related Work}
The ease in which static data about malware can be obfuscated and the extent to which that increases the limits of static analysis approaches to malware detection have been extensively evaluated \cite{moser2007limits}. Intrusion detection using approaches ranging from variants of SVMs and neural networks using data pertaining to network traffic have been attempted with significant success \cite{mukkamala2002intrusion}\cite{thaseen2017intrusion}\cite{cannady1998artificial}\cite{ibrahim2010anomaly}. Specifically in analysis of system calls, there has been implementations on Android phones \cite{malik2016system}\cite{chaba2017malware}. In application to Windows operating systems, there has been an approach using n-grams of Windows API calls and SVM, a further discussion on n-grams of system calls and its variants dubbed Sequencegrams, and a paper assessing the use of one, two, three, and four grams of system calls weighted by TF-IDF in malware detection \cite{veeramani2012windows}\cite{hubballi2011sequencegram}\cite{canzanese2015detection}.

This study aims to build upon the related work by attempting and comparing different approaches in using document classification techniques on system call analysis in detecting malware. It seeks to prove that certain choices are noticeably useful in this approach. It also proposes a system that integrates this approach of using system calls for malware detection into a deployable form.

\section{Procedure}
\label{sec:examples}
Note: The feature extraction process and the detection algorithms were implemented using the Scikit-learn  libary \cite{scikit-learn}.
\subsection{Dataset Preparation}

Various malware corpora were collected from online sources such as VirusShare and The Zoo \cite{virusshare}\cite{thezoo}. VirusTotal, a website that displays results of testing the program on various antivirus software, was used to validate whether or not a given program was malicious or benign \cite{virustotal}. A program was deemed to be malicious if more than 80\% of the antivirus softwares shown in VirusTotal deemed it to be malicious, and a program was deemed to be benign if all agreed that it was harmless.

NtTrace, a native API tracer for Windows, was used to collect system call traces. \cite{orr}. This program is designed to run in the command prompt, specifying the path to the program as well as options on how it should be traced (filters, logging only errors, etc). As running malware on personal computers will not be safe for various reasons, a virtual machine was created to be used as a host. VirtualBox was used for this purpose \cite{virtualbox}. Malware samples were executed on a Windows operating system hosted in VirtualBox. Batch scripts were used to automate the process of collecting the system calls. A batch script is a file that executes a series of commands \cite{laurie}. Batch scripts can run these commands in loops, automating the process of tracing the system calls executed by hundreds of programs.

\subsection{Text Preprocessing}

Because the features considered are sequences of system call functions, for the purpose of this research, the parameters were not considered as a feature. A script was used to process the system call logs generated by NtTrace to remove the parameters, only leaving the function names. Furthermore, sections of the logs that were not related to system call function names, such as logs informing unloading of DLLs, were removed as well. FIG. 2 demonstrates this pre-processing process. 

\begin{figure}
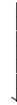

\fbox{
    \parbox{24em}{
    \texttt{
\begin{flushleft} Unload of DLL at 04ED0000\newline
Unload of DLL at 04FC0000\newline
NtQueryPerformanceCounter( Counter=0x4e9f9c8 [3.01683e+009], Freq=null ) => 0 \newline
NtProtectVirtualMemory( ProcessHandle=-1, BaseAddress=0x4e9f9f4 [0x77eae000], Size=0x4e9f9f8\newline
\end{flushleft}
	  }
 	 }
     }
 $
\centering
\newline\newline\newline
\tab \tab \xdownarrow{0.8cm}
\newline\newline\newline
$
\fbox{
    \parbox{15em}{
    \texttt{
    \begin{flushleft}
NtQueryPerformanceCounter\newline
NtProtectVirtualMemory
\end{flushleft}
  }
  }
  }

\caption{\label{fig:your-figure} Processing Trace}

\end{figure}

% Commands to include a figure:
%\begin{figure}
%\includegraphics[width=\textwidth]{your-figure's-file-name}
%\caption{\label{fig:your-figure}Caption goes here.}
%\end{figure}

\renewcommand{\thetable}{\arabic{table}}

\subsection{Feature Extraction}
We define some terms and variables that will be referred to as following: \newline
\newline
$D=\{d_1, d_2, \ldots, d_n\}$  \newline\newline
$V=\{w_1, w_2, \ldots, w_n\}$ \newline\newline
$f_d(w)=$ frequency of the word $w{\in}V$ in $d{\in}D$ \newline\newline
$\vec{t_d}=(f_d(w_1), f_d(w_2), \ldots, f_d(w_{\lvert v| \lvert} ))$ \newline\newline
$D$ is the document corpus and $d$ represents a particular document. \newline\newline
$V$ is the vocabulary and $w$ represent each word that appears in the corpora\newline\newline

\begin{bf}
1. Bags of Words Model
\newline 
\end{bf}

Bags of words is the way in which features are extracted from text to be used for the machine learning algorithms. The idea behind ``bag'' is that order is not accounted for; this model only takes into account whether certain words occur in a document, not where they occur in a document \cite{brownlee_2017}. In this research, this model is applied in the sense that the ``words'' are system calls and the ``documents'' are logs of system calls. In this model, set $V$ is built as the vocabulary set of all unique system calls (represented by $w_n$, where $n$ is the index of the system call in the vocabulary). Each document is represented by how many times every $w_n$ occurs in the document, as represented by $\vec{t_d}$, where $f_d(w_n)$ represents the frequency of the system call of index $n$ in the log $d$.
\newline

\begin{bf}
2. $N$-Grams Model
\newline 
\end{bf}

This approach to use single system calls as features is a unigram approach, as the features consist of just one ``gram''. This falls short in many text classification tasks as the bags-of-words model does not take into account sequences of text. The $n$-gram model accommodates for this issue, where the features are sequences of system calls instead of a single system call, therefore analyzing frequencies of sequences of system calls instead of frequencies of individual system calls. In an $n$-grams model, the features are sequences of n number of system calls. FIG. 3 demonstrates the conversion of unigrams to bigrams.

\begin{figure}
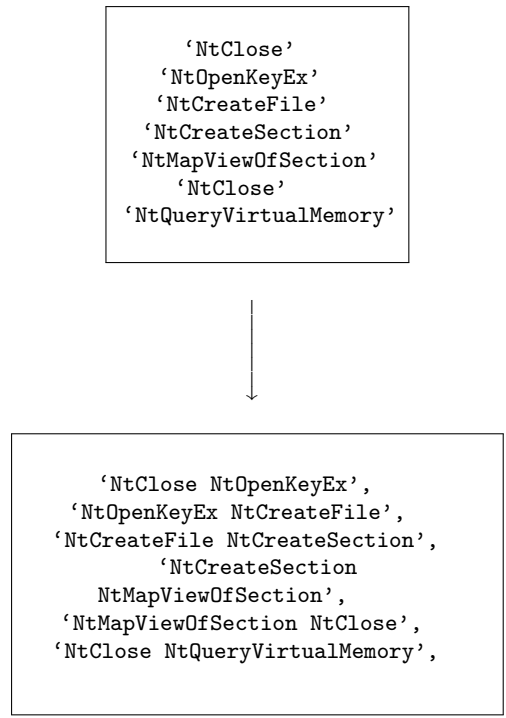

\fbox{
    \parbox{0.2\textwidth}{
\begin{center}
\texttt{
`NtClose'\newline
`NtOpenKeyEx' \newline
`NtCreateFile' \newline
`NtCreateSection'\newline
`NtMapViewOfSection'\newline
`NtClose'\newline
`NtQueryVirtualMemory'
	  }
	 \end{center}
	  }
 	 }
     
 $
\centering
\newline\newline
\tab \tab \xdownarrow{0.8cm}
\newline\newline\newline
$
\fbox{
    \parbox{0.34\textwidth}{
    \texttt{
\begin{center}
\texttt{
`NtClose NtOpenKeyEx',\newline
`NtOpenKeyEx NtCreateFile',\newline
`NtCreateFile NtCreateSection',\newline
`NtCreateSection NtMapViewOfSection',\newline
`NtMapViewOfSection NtClose',\newline
`NtClose NtQueryVirtualMemory',\newline
  }
  \end{center}
  }
  }
  }

\caption{\label{fig:your-figure}  Unigrams to Bigrams Conversion}

\end{figure}

In this research, the implemented $n$-grams were eight-grams, nine-grams, and ten-grams. This was derived through grid search algorithms specified later that showed that this combination provided the best accuracy. This was expected, considering that operations of any program would typically require reasonably long sequences of system calls.
\newline \newline
\begin{bf}
3. Term Frequency - Inverse Document Frequency (TF-IDF) Weighting
\newline 
\end{bf}

In any document corpus, certain words are more common than others, such as ``the'', ``a'', ``of'', and more. An assumption was made that the same idea applies to logs of system calls, that certain sequences pertain to all operations of any program in an operating system and certain sequences pertain to specific operations that may or may not be malicious. In document classification, one of the most popular model to derive weights to terms that occur in a document is the TF-IDF model. 

Term Frequency (TF) refers to the number of times a certain word occurs in a document. Inverse Document Frequency (IDF) refers to the amount of times the word occurs throughout the corpus.
The TF-IDF weight of a term is computed as following:
\newline\newline
$tf(w,d)=f_d(w)$: frequency of $w$ in document $d$ \newline
$idf(w, D)=log\frac{1+\lvert D \lvert}{1+df(d,w)}$\newline\newline
Where $df(d,w)$ is the number of documents the word $w$ appears in.\\

This is a logarithmically scaled value of the number of documents in the corpus divided by the number of times word w appears throughout the corpus. \newline\newline
$tfidf(w,d,D)=tf(w,d) \times idf(w,D)$ \newline\newline
The $tfidf$ value increases proporationally by the frequency of $w$ in a document, decreases proportionally by the $log$ of the frequency of $w$ in the corpus. The assumption is that a word that is more prevalent throughout the corpus is more likely to be less significant.\\

The resulting vectors comprised of raw $tfidf$ values that represent each document are normalized using the Euclidean norm:
\begin{center}
$v_{norm} = \frac{v}{\sqrt{v_1^2 + v_2^2+...+v_n^2}}$
\end{center}
For instance: \newline\newline
$tfidf_{raw} = [10,3,1]$\newline\newline
$tfidf_{normalized} = \frac{[10,3,1]}{\sqrt{10^2 + 3^2 + 1^2}} = [0.953,0.286,0.095]$

\subsection{Detection Algorithm}
\begin{bf}
1. Support Vector Machines (SVM)
\newline 
\end{bf}

The objective of a SVM classifier is to learn a decision boundary hyperplane that optimally separates the dataset. The optimized decision boundary is then used to compute whether or not a new data point that it is tested on pertains to malware or not.

Given a training data, $(x_i,y_i)$, where $x_i$ is the collection of features (TF-IDF values of n-grams of system calls) of a document and $y_i{\in}\{-1,1\}$, the objective of SVM is to learn a classifier $f(x)$ so that: 
\[   
f(x_i) = 
     \begin{cases}
       \geqslant0 & y_i=+1 \\
       \leqslant0 & y_i=-1
     \end{cases}
\]
Where $f(x_i)$ is defined by:
\begin{center}
$f(x_i)=w^T x+b$\\
\end{center}

The best decision boundary is determined by the margins between the decision boundary and the support vectors, the data points closest to the decision boundary. The best decision boundary is defined by one that has the largest margins from the support vectors.\newline\newline
\begin{bf}
2. Stochastic Gradient Descent (SGD) Optimization Using Various Regularization Terms
\newline 
\end{bf}

The Stochastic Gradient Descent algorithm optimizes this optimal decision boundary. With the goal of learning  $f(x_i)=w^T x+b$, the best model parameters $w{\in}R^m$ is computed by minimizing the regularized training error, which is given by:
\begin{center}
$
E(w,b)=\frac{1}{n}\sum\limits_{i=1}^n L(f(x_i),y_i)+{\alpha}R(w)
$
\end{center}
Where $L$ is the loss (cost) function, which measures the error of the model, and $R$ is a regularization term, reducing the likelihood that the model is overfitting. The hinge loss function ($L$) is defined by the following:

\begin{center}
$
L(y,\hat{y})=max(0,1-y\times\hat{y})
$
\end{center}
where $y$ is the prediction of the classifier and ${\hat{y}}{\in}\{-1,1\}$ is the intended output.

The regularization terms $R(w)$ implemented were L1, L2, and Elastic Net penalties.

\begin{center}
\begin{bf}L1 Penalty:
$R(w):=\frac{1}{2} \sum\limits_{i=1}^n\lvert w_i \lvert$\\
\bf{L2 Penalty:}
$R(w):=\frac{1}{2} \sum\limits_{i=1}^nw_i^2$\newline \\ 
\bf{Elastic-Net Penalty:}
$R(w):=\frac{\varphi}{2} \sum\limits_{i=1}^nw_i^2+(1-\varphi)\sum\limits_{i=1}^n\lvert w \lvert$\\
\end{bf}
A convex combination of L1 and L2 Penalty.
\end{center}
Given this error function, the $w$ parameter is updated iteratively by the following operation:
\begin{center}
$w{\leftarrow}w-\eta(\alpha\frac{\partial R(w)}{\partial w}+\frac{\partial L(w^Tx+b,y_i)}{\partial w})$
\end{center}
Where $\eta$ is the learning rate that controls the step size:
\begin{center}
$\eta^{(t)}=\frac{1}{\alpha(t_0+t)}$
\end{center}
Where $t$ is the time step.\newline\newline

\begin{bf}
3.Coordinate Descent Method Using LibLinear's Linear Support Vector Machine
\end{bf}\newline\newline
Another SVM was optimized using a different framework, which was a Linear SVM implemented using LibLinear, developed by the machine learning group in the National Taiwan University. LibLinear uses the Coordinate Descent method to optimize the linear support vector machines.

The width of the margin of a SVM can be represented by: $\frac{2}{\lvert \lvert \vec{w} \lvert \lvert}$, and the objective is: $y_i(\vec{x_i}\vec{w}+b)\geqslant1$. This poses a constrained optimization problem:

\begin{center}
$
 \min \frac{1}{2} \lvert\lvert\vec{w} \lvert \lvert ^2$ where $ y_i \vec{x_i}\vec{w}+b\geqslant1
$
\end{center}
The Lagrange form of which is:

\begin{center}
$
L(\vec{w},b, \alpha) = \frac{1}{2} \lvert\lvert\vec{w} \lvert \lvert ^2 - \sum \alpha_i  [y_i(\vec{w} \vec{x_i}+b)-1]
$
\end{center}

and because:
\begin{center}
$\frac{\partial L}{\partial \vec{w}} = \vec{w} - \sum\alpha_iy_ix_i=0$\\
$\frac{\partial L}{\partial b} = - \sum\alpha_iy_i=0$
\end{center}
So:
\begin{center}
$\vec{w} = \sum\alpha_iy_ix_i$\\
$\sum\alpha_iy_i=0$
\end{center}

The Lagrangian can be simplified to (due to the properties of the partial derivatives) the Wolfe dual form:
\begin{center}
$
L(\vec{w},b, \alpha) = \sum \alpha_i - \frac{1}{2}\sum_i\sum_j\alpha_i\alpha_jy_iy_jx_i^Tx_j
$
\end{center}

Minimizing $L(\vec{w},b, \alpha)$ is to:

\begin{center}
$
\min_\alpha f(\alpha) = \frac{1}{2}\alpha^TQ\alpha-e^T\alpha
$
\end{center}

Where Q is a matrix where $Q_{ij} = y_iy_jx_i^Tx_j$ and $e$ is a vector of all ones.\newline

The Coordinate Descent algorithm finds the minimum of a multivariate function F(x) by solving univariate optimization problems iteratively through all its variables (inner iteration), and iteratively doing this several times (outer iteration). With the following definition, $k$ being the outer iteration and $i$ being the inner iteration:

\begin{center}
$a^{k,i} = [a_1^{k+1}, ..., a^{k+1}_{i-1},...,a^k_i,...,a^k_l]^T$
\end{center}
solving the following univariate function:
\begin{center}
$\min_d f(\alpha^{k,i}+de_i)$
\end{center}
Where $e_i=[0,...,0,1,0,...,0]^T$ and $d$ is the scaler to it, representing a step towards that direction. It becomes apparent that the minimum is found when $d = 0$, where there is no where to move to minimize the function $f$, which is when 
\begin{center}
$\nabla^P_if(\alpha^{k,i})=0$
\end{center}
Where $\nabla^Pf(\alpha)$ refers to the projected gradient.\newline\newline

\begin{bf}
4. Hyperparameter Optimization
\newline 
\end{bf}

Machine learning models require varying constraints, learning rate, and more that can affect the performance of the model. These factors are called hyperparameters and they must be tuned for a model to produce optimal performance. The method we implemented was Grid Search, which demonstrated by Algorithm 1.\newline

\begin{algorithm}[H]
  \caption{Grid Search}
  \label{EPSA}
   \begin{algorithmic}[1]
   \State {$alpha_{grid} = [100, 10, 1, 0.1, ... , 0.0000001]$\\
$tol_{grid} = [100, 10, 1, 0.1, ... , 0.00005]$\\
$fl$-$score = []$}
  \For{\texttt{$\alpha$ in $alpha_{grid}$}}
 	 \For{\texttt{$tol$ in $tol_{grid}$}}
       		\State \texttt{model.train(alpha=$\alpha$, tol=tol)}
		\State \texttt{pred = model.predict()}
		\State \texttt{flscore = evaluate(pred, $y$-$true$)}
		\State \texttt{$fl$-$score$.append(flscore)}
  	\EndFor
  \EndFor
  \State \texttt{outputMaxValue($fl$-$score$)}
   \end{algorithmic}
\end{algorithm}

The benefits of this process is visualized by Figure 4. Darker shades of red indicates a higher value in fl\_score of the model with the corresponding combination.

\begin{figure}
\includegraphics[scale=0.4]{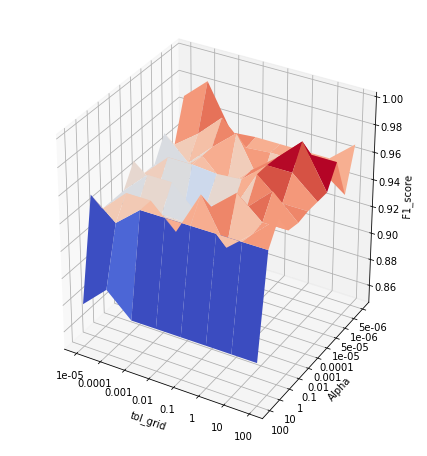}
\caption{\label{fig:images/grid_search}Grid Search Output Visualized}
\end{figure}

\subsection{Testing Procedure}
\begin{bf}
1. Splitting Training \& Testing Set
\newline
\end{bf}

In order to create a realistic scenario for testing, the corpus of data was split into a training set which the classifiers were trained on, and a testing set which the classifiers have not seen before. This way the results of the classifiers could be trusted, since any system call logs encountered in the testing data would not have been encountered by the classifiers, and would not be able to directly classify. 
\newline\newline\newline
\begin{bf}
2. Precision Score
\newline
\end{bf}

The precision score is the number of correctly identified malicious programs over the number of true positives plus false positives:
\newline
\begin{center}
$Ps=\frac{tp}{tp+fp}$\\
\end{center}

This provides a practical score to judge the performance of each classifier, as false positives can be as troublesome as false negatives in certain cases, and are more noticeable during operation by users. Considering that a lopsided proportion of malicious traces may skew the accuracy if it only accounts for true positives, this precision score displays a better idea of the accuracy of the model.
\newline\newline\newline
\begin{bf}
3. Recall Score
\newline
\end{bf}

The recall score is the number of true positives divided by the number of true positives plus false negatives: 
\begin{center}
$Rs=\frac{tp}{tp+fn}$
\end{center}
This gives an intuitive rate of how many malicious programs were detected.
\newline\newline
\begin{bf}
4. F1 Score
\newline
\end{bf}

The F1 score is a weighted average of the precision score and recall score:

\begin{center}
$F1=2\frac{Ps*Rs}{Ps+Rs}$
\end{center}
It offers a different view of the accuracy of each classifier, in which false positives and false negatives are both integrated. This acts as a more holistic means of comparison.

\maketitle

\section{Results}
The dataset was divided by a 1:4 ratio; 80\% were used to train the classifiers and 20\% were used to test the classifiers. Also, the proportion of malicious trace in the testing set was 63.7\%, and judgement on the scores provided below need to take into account this proportion. 
Table 1 shows the results of primal SVM optimized using Stochastic Gradient Descent and Table 2 shows the results of using LibLinear to optimize the dual form of SVM using the Coordinate Descent method.

\begin{table}[H]
	\begin{center}
    \begin{tabular}{|l|l|l|l|}
    \hline
    ~             & Precision & Recall & F1-Score \\ \hline
    Benign        & 1.00     & 0.82  & 0.92     \\ \hline
    Malware       & 0.94      & 1.00   & 0.97     \\ \hline
    Average/Total & 0.96      & 0.95   & 0.95     \\ \hline
    \end{tabular}
    \end{center}
\end{table}
\captionof{table}{Results of SGD Classifier}

\begin{table}[H]
	\begin{center}
    \begin{tabular}{|l|l|l|l|}
    \hline
    ~             & Precision & Recall & F1-Score \\ \hline
    Benign        & 1.00     & 0.79  & 0.88     \\ \hline
    Malware       & 0.91      & 1.00   & 0.95     \\ \hline
    Average/Total & 0.94      & 0.93   & 0.93     \\ \hline
    \end{tabular}
    \end{center}
\end{table}
\captionof{table}{Results of LibLinear}

It is crucial for a malware detection program to not only detect malware but have a very low false positive rate. The programs were optimized accordingly, preferring a better score in its precision score for benign data than for malware.\newline\newline
This TPR (True Positive Rate) vs. FPR (False Positive Rate) tradeoff can be represented through the ROC (Receiver Operating Characteristic) curve, which plots the values of TPR and FPR at different decision thresholds, shown in FIG. 5 for the SGD Classifier. FIG. 6 is the ROC curve for the SVM implemented using LibLinear.\\

\begin{figure}
\includegraphics[scale=0.3]{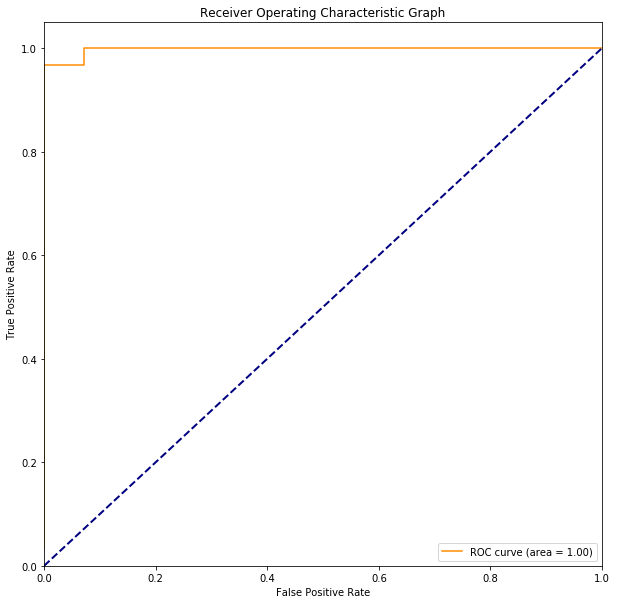}
\caption{\label{fig:rocsgd}ROC Curve of SGDClassifier (AUC=1.00)}
\end{figure}

\begin{figure}
\includegraphics[scale=0.3]{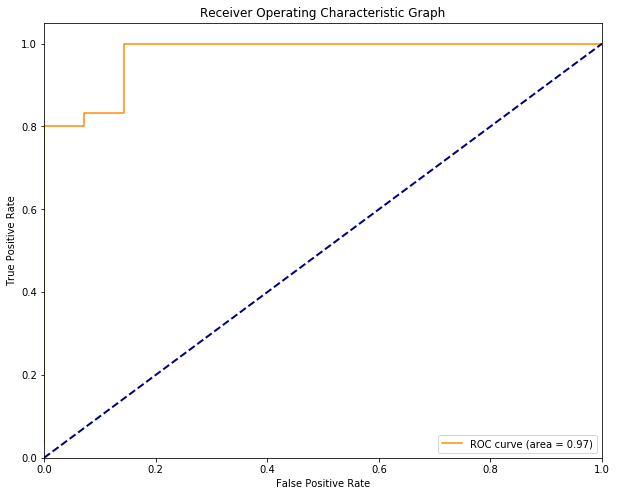}
\caption{\label{fig:rocsvc}ROC Curve of LibLinear Classifier (AUC=0.97)}
\end{figure}

	Furthermore, refer to Table 2 to compare the effectiveness of the options in the feature extraction process such as TF-IDF weighting and the use of n-grams. The average values were computed by calculating the mean of the respective scores of the classifiers listed previously on Figure 4.\newline

\begin{table}[H]
    \begin{tabular}{|l|l|l|l|}
    \hline
    ~             & Avg. Precision & Avg. Recall \\ \hline
    TFIDF \& 10-Grams        & 0.93      & 0.92  \\ \hline
    TFIDF \& Unigrams       & 0.90      & 0.89    \\ \hline
    Term Frequency \& 10-grams  & 0.78      & 0.78      \\ \hline
    None & 0.75 & 0.72 \\ \hline
    \end{tabular}
\end{table}
\captionof{table}{Average Scores Comparing Options}

This is a valid testimony to the effectiveness of these options not only in document classification tasks but also in applying document classification on system call traces. \\
Furthermore, it is also observed that SVM classifiers were able to classify the traces in the testing set in the shortest time, which is one of the reasons why SVMs are often preferred in document classification tasks. Table 3 shows their training and testing time in comparison to other classifiers.

\begin{table}[H]
    \begin{tabular}{|l|l|l|l|}
    \hline
    ~             & Training Time (s) & Testing Time (s) \\ \hline
    L2 Penalty SVM SGD   & 0.098s      & 0.000s  \\ \hline
    Passive-Aggressive & 0.112s & 0.000s \\ \hline
    kNN      & 0.003s      & 0.059s    \\ \hline
    Bernoulli Naive Bayes & 0.024s      & 0.018s      \\ \hline
    \end{tabular}
\end{table}
\captionof{table}{Training \& Testing Time Comparison}

Although these are small absolute differences, as the classification task scales, these differences may amplify to produce a considerable difference.

\maketitle
\section{Conclusions \& Contributions}

This paper demonstrates the effectiveness of applying document classification techniques on system call traces for the purpose of detecting malicious programs. The effectiveness and time complexities of certain algorithms as well as certain options popular in the realm of document classification was compared to judge the validity of these options.

\indent

The classifiers trained in this research were also able to provide features (TF-IDF values of 10-grams of system calls) that they deemed to be the most significant. These system call sequences could be avenues for further research to determine the nature of these sequences.\newline
The sequences in Table 4 were produced by the SVM classifier optimized by SGD using L1 regularization term:

\begin{table}[h]
{\renewcommand\arraystretch{2}
\begin{tabular}{|l|l|l|} \hline
7.20687749597& \multicolumn{2}{p{6cm}|}{ ntqueryinformationthread ntqueryinformationthread ntqueryinformationthread ntqueryinformationthread ntqueryinformationthread ntqueryinformationthread ntqueryinformationthread ntqueryinformationthread ntqueryinformationthread ntqueryinformationthread} \\ \hline

6.41295595185& \multicolumn{2}{p{6cm}|}{ ntmapviewofsection ntunmapviewofsection ntmapviewofsection ntunmapviewofsection ntmapviewofsection ntunmapviewofsection ntmapviewofsection ntunmapviewofsection ntmapviewofsection ntunmapviewofsection } \\ \hline

4.75759889433& \multicolumn{2}{p{6cm}|}{ ntsetinformationfile ntreadfile ntsetinformationfile ntreadfile ntsetinformationfile ntreadfile ntsetinformationfile ntreadfile ntsetinformationfile ntreadfile } \\ \hline

4.75759889433& \multicolumn{2}{p{6cm}|}{ ntreadfile ntsetinformationfile ntreadfile ntsetinformationfile ntreadfile ntsetinformationfile ntreadfile ntsetinformationfile ntreadfile ntsetinformationfile } \\ \hline
\end{tabular}}
\captionof{table}{SVM SGD L1 Most Informative Features}

\end{table}

Greater values of coefficients indicate greater relevance for the specific classifier. These features were the few among 237588 sequences that these classifiers took into consideration in classifying these logs of system calls. Table 5 shows the sequences produced by the same classifier as that of Table 4, but using L2 Penalty.\newline

\begin{table}[h]
{\renewcommand\arraystretch{2}
\begin{tabular}{|l|l|l|} \hline
3.0649184842& \multicolumn{2}{p{6cm}|}{ntdelayexecution ntdelayexecution ntdelayexecution ntdelayexecution ntdelayexecution ntdelayexecution ntdelayexecution ntdelayexecution ntdelayexecution ntdelayexecution
} \\ \hline
2.07007999086& \multicolumn{2}{p{6cm}|}{ ntgetcurrentprocessornumber ntgetcurrentprocessornumber ntalpcsendwaitreceiveport ntgetcurrentprocessornumber ntgetcurrentprocessornumber ntalpcsendwaitreceiveport ntgetcurrentprocessornumber ntgetcurrentprocessornumber ntalpcsendwaitreceiveport ntgetcurrentprocessornumber
} \\ \hline
1.55237596748& \multicolumn{2}{p{6cm}|}{ntdeviceiocontrolfile ntclose ntcreateevent ntdeviceiocontrolfile ntclose ntcreateevent ntdeviceiocontrolfile ntclose ntcreateevent ntdeviceiocontrolfile
 } \\ \hline
1.52963083335& \multicolumn{2}{p{6cm}|}{ntclose ntcreateevent ntdeviceiocontrolfile ntclose ntcreateevent ntdeviceiocontrolfile ntclose ntcreateevent ntdeviceiocontrolfile ntclose
} \\ \hline
\end{tabular}}
\captionof{table}{SVM SGD L2 Most Informative Features}

\end{table}

NtMalDetect (https://github.com/codeandproduce\\
/NtMalDetect) is an open source project that utilizes the classifiers trained from this research to put them into an executable form. It utilizes boosted classifiers, combining inputs of various classifiers to produce one output, to detect malicious program before and after execution. 

\section{Acknowledgements}
Thank you staff members and colleagues at Shanghai American School Puxi Campus and members of Coderbunker for providing the resources and guidance to be able to conduct this research and enter this project to the Intel International Science and Engineering Fair. This project has been recognized by being awarded the Yale Science and Engineering Award and being named a finalist project to the Intel International Science and Engineering Fair (ISEF). At Intel ISEF, this project was recognized by receiving special awards from  King Abdulaziz \& His Companions Foundation for Giftedness and Creativity, "MAWHIBA" and from China Association for Science and Technology. It was awarded \$1200 and \$1000 from those organizations, respectively. At the fair, this project received the 4th place Grand Award for the Systems Software category, awarding it \$500.
\def\bibsection{\section*{\refname}} 
\bibliography{references}

\begin{thebibliography}{10}

\bibitem{brownlee_2017}
Jason Brownlee.
\newblock A gentle introduction to the bag-of-words model, Nov 2017.
\newblock URL:
  \url{https://machinelearningmastery.com/gentle-introduction-bag-words-model/}.

\bibitem{cannady1998artificial}
James Cannady.
\newblock Artificial neural networks for misuse detection.
\newblock In {\em National information systems security conference}, volume~26.
  Baltimore, 1998.

\bibitem{canzanese2015detection}
Raymond~J Canzanese~Jr.
\newblock {\em Detection and classification of malicious processes using system
  call analysis}.
\newblock Drexel University, 2015.

\bibitem{chaba2017malware}
Sanya Chaba, Rahul Kumar, Rohan Pant, and Mayank Dave.
\newblock Malware detection approach for android systems using system call
  logs.
\newblock {\em arXiv preprint arXiv:1709.08805}, 2017.

\bibitem{distler2007malware}
Dennis Distler and Charles Hornat.
\newblock Malware analysis: An introduction.
\newblock {\em SANS Institute InfoSec Reading Room}, pages 18--19, 2007.

\bibitem{ghaffari}
Parsa Ghaffari.
\newblock Text analysis 101: Document classification.
\newblock URL:
  \url{https://www.kdnuggets.com/2015/01/text-analysis-101-document-classification.html}.

\bibitem{hubballi2011sequencegram}
Neminath Hubballi, Santosh Biswas, and Sukumar Nandi.
\newblock Sequencegram: n-gram modeling of system calls for program based
  anomaly detection.
\newblock In {\em Communication Systems and Networks (COMSNETS), 2011 Third
  International Conference on}, pages 1--10. IEEE, 2011.

\bibitem{ibrahim2010anomaly}
Laheeb~Mohammad Ibrahim.
\newblock Anomaly network intrusion detection system based on distributed
  time-delay neural network (dtdnn).
\newblock {\em Journal of Engineering Science and Technology}, 5(4):457--471,
  2010.

\bibitem{laurie}
Vic Laurie.
\newblock Batch files (scripts) in windows.
\newblock URL: \url{https://commandwindows.com/batch.htm}.

\bibitem{malik2016system}
Sapna Malik and Kiran Khatter.
\newblock System call analysis of android malware families.
\newblock {\em Indian Journal of Science and Technology}, 9(21), 2016.

\bibitem{metz_2017}
Cade Metz.
\newblock Google says its ai catches 99.9 percent of gmail spam, Jun 2017.
\newblock URL:
  \url{http://www.wired.com/2015/07/google-says-ai-catches-99-9-percent-gmail-spam}.

\bibitem{moser2007limits}
Andreas Moser, Christopher Kruegel, and Engin Kirda.
\newblock Limits of static analysis for malware detection.
\newblock In {\em Computer security applications conference, 2007. ACSAC 2007.
  Twenty-third annual}, pages 421--430. IEEE, 2007.

\bibitem{mukkamala2002intrusion}
Srinivas Mukkamala, Guadalupe Janoski, and Andrew Sung.
\newblock Intrusion detection using neural networks and support vector
  machines.
\newblock In {\em Neural Networks, 2002. IJCNN'02. Proceedings of the 2002
  International Joint Conference on}, volume~2, pages 1702--1707. IEEE, 2002.

\bibitem{thezoo}
Yuval Nativ.
\newblock The zoo.
\newblock URL: \url{http://thezoo.morirt.com/}.

\bibitem{orr}
Roger Orr.
\newblock Nttrace.
\newblock URL: \url{http://rogerorr.github.io/NtTrace/}.

\bibitem{scikit-learn}
F.~Pedregosa, G.~Varoquaux, A.~Gramfort, V.~Michel, B.~Thirion, O.~Grisel,
  M.~Blondel, P.~Prettenhofer, R.~Weiss, V.~Dubourg, J.~Vanderplas, A.~Passos,
  D.~Cournapeau, M.~Brucher, M.~Perrot, and E.~Duchesnay.
\newblock Scikit-learn: Machine learning in {P}ython.
\newblock {\em Journal of Machine Learning Research}, 12:2825--2830, 2011.

\bibitem{thaseen2017intrusion}
Ikram~Sumaiya Thaseen and Cherukuri~Aswani Kumar.
\newblock Intrusion detection model using fusion of chi-square feature
  selection and multi class svm.
\newblock {\em Journal of King Saud University-Computer and Information
  Sciences}, 29(4):462--472, 2017.

\bibitem{veeramani2012windows}
R~Veeramani and Nitin Rai.
\newblock Windows api based malware detection and framework analysis.
\newblock In {\em International conference on networks and cyber security},
  volume~25, 2012.

\bibitem{virtualbox}
{VirtualBox}.
\newblock Virtualbox.
\newblock URL: \url{https://www.virtualbox.org/}.

\bibitem{virusshare}
VirusShare.
\newblock Virusshare.
\newblock URL: \url{https://virusshare.com/}.

\bibitem{virustotal}
VirusTotal.
\newblock Virustotal.
\newblock URL: \url{https://www.virustotal.com/}.

\end{thebibliography}
 \bibliographystyle{plainurl}

\end{document}